\documentclass{ws-ijmpd}
\usepackage{amsmath,amssymb,amsfonts}
\usepackage{latexsym}
\usepackage{revsymb}
\usepackage{graphicx}
\usepackage{hyperref}
\usepackage[compress]{cite}

\begin{document}
\title{Observable Effects of Quantum Gravity}

\author{Lay Nam Chang${}^*$,
Djordje Minic${}^\dagger$, 
Chen Sun${}^\ddagger$,
and
Tatsu Takeuchi${}^\S$
}

\address{
Center for Neutrino Physics, Department of Physics, Virginia Tech,
Blacksburg, VA 24061 USA\\
${}^*$laynam@vt.edu,
${}^\dagger$dminic@vt.edu,
${}^\ddagger$sunchen@vt.edu,
${}^\S$takeuchi@vt.edu
}

\maketitle

\begin{abstract}
We discuss the generic phenomenology of quantum gravity and, in particular, argue that the observable effects of quantum gravity, associated with new, extended, non-local, non-particle-like quanta,  and
accompanied by a dynamical energy-momentum space, are not necessarily Planckian and that they could be observed at much lower and experimentally accessible energy scales.
\\
\\
Essay written for the Gravity Research Foundation 2016 Awards for Essays on Gravitation.
\end{abstract}

\setcounter{footnote}{0}
\renewcommand{\thefootnote}{\alph{footnote}}

\section{The Quest for Quantum Gravity}

The two main legacies of 20th century physics are relativity 
and quantum theory.
Both frameworks have been tested in numerous experiments, including the recent detection of gravitational waves by LIGO \cite{Abbott:2016blz}, and the completion of the fundamental
quantum field theoretic framework of the Standard Model (SM) of particle physics, exemplified by the LHC discovery of the Higgs particle \cite{Aad:2012tfa,Chatrchyan:2012xdj}. 
However, the conceptual foundations of general relativity and quantum theory still stand apart.
It is thus often stated that one of the most outstanding problems of physics is to find a unified habitat for these two
great frameworks. That is the problem of quantum gravity \cite{Kiefer:2012boa}.

In this essay, we address the question of  the generic phenomenological effects of quantum gravity and especially the crucial role played by the minimal length. 
In particular, we argue that the observable effects of quantum gravity may not necessarily be Planckian ($\sim\! 10^{19}\,\mathrm{GeV}$) and that the essential phenomenology of quantum gravity may be observable at much lower, and experimentally accessible, energy scales.

\section{Quantum Gravity and the Minimal Length}

We start our discussion with the concept of a minimal length, a feature expected of any candidate theory of quantum gravity,
since gravity itself is characterized by the Planck scale $\ell_P = \sqrt{\hbar G_N/c^3} \sim 10^{-35}\,\mathrm{m}$. 
One can re-analyze the Heisenberg microscope gedanken experiment in the presence of gravity, and
claim that the following minimal length uncertainty relation (MLUR) should hold \cite{Garay:1994en,Chang:2011jj,Hossenfelder:2012jw}:
\begin{equation}
\delta x \;\sim\; \left( \frac{\hbar}{\delta p} \,+\, \ell_P^2\,\frac{\delta p}{\hbar} \right)\;.
\label{MLUR1}
\end{equation}
This relation, which implies that $\delta x$ is bounded from below by $\ell_P$,
can be argued on many different grounds, in canonical gravity, in string theory (via T-duality), etc. and it can be intuitively understood using simple dimensional analysis.
The actual existence of the minimal length can, in principle, be tested with LIGO-like laser interferometry, which represents the best existing probe of the fine structure of space and time \cite{Chen:2013sgs}.

It is sometimes stated that the MLUR is a probe-dependent statement, and thus that it does not carry any universal and probe-independent information.
In the following we wish to argue the opposite: the minimal length leads to some generic predictions about the low energy phenomenology of quantum gravity even in the absence of the universally accepted and final version of quantum gravity.

First we rewrite the above relation more precisely as
\begin{equation}
\delta x\,\delta p \;\ge\; \dfrac{\hbar}{2}\left(1 + \beta\, {\delta p}^2\right)\;,\qquad
\beta \;=\; \ell_P^2/\hbar^2\;,
\label{MLUR2}
\end{equation}
which can be reproduced within Quantum Mechanics (QM) by a deformation of the canonical Heisenberg algebra:
\begin{equation}
\dfrac{1}{i\hbar}[\,\hat{x},\,\hat{p}\,] \;=\; 1 
\quad\rightarrow\quad
\dfrac{1}{i\hbar}[\,\hat{x},\,\hat{p}\,] \;=\; 1+\beta\hat{p}^2\;.
\end{equation}
Though non-relativistic, such a deformation can be expected as a consequence
of quantum gravity in some limit if the MLUR is to hold.
The multi-dimensional generalization of this deformation, though constrained by the Jacobi identity, is not unique.
While quantum gravity should eventually dictate what the deformation should be,
we find the following particularly attractive \cite{Chang:2016sae}:
\begin{eqnarray}
[\,\hat{p}_i,\,\hat{p}_j\,]  =  0 \vphantom{\Big|},
\quad
\dfrac{1}{i\hbar}[\,\hat{x}_i,\,\hat{p}_j\,]
 =  (1-\beta\hat{\mathbf{p}}^2)\delta_{ij}+2\beta\,\hat{p}_i\hat{p}_j,
\quad
\dfrac{1}{i\hbar}[\,\hat{x}_i,\,\hat{x}_j\,]
=  4\beta\hat{L}_{ij}\;.
\label{MLUR-CommutationRelations}
\end{eqnarray}
Here, $\hat{L}_{ij} = (\hat{x}_i\hat{p}_i-\hat{x}_j\hat{p}_i)/(1-\beta\mathbf{p}^2)$ is the generator of rotations,
\textit{i.e.} the angular momentum operator, which always appears in the expression for $[\,\hat{x}_i,\,\hat{x}_j\,]$.
In these relations, our choice of $[\,\hat{x}_i,\,\hat{p}_j\,]$ has rendered the coefficient of $\hat{L}_{ij}$ a constant,
simplifying the algebra considerably.
We note that the non-commutativity of the coordinate operators is to be expected in quantum gravity
in which the geometry of spacetime itself is quantized.
The $\hat{x}_i$ operators which satisfy these commutation relations 
in $D$-dimensions can be realized as rotation/boost operators in $(D+1)$-dimensions, 
while the $\hat{p}_i$ operators can be realized as operators obtained by stereographic projection of a
$(D+1)$-dimensional hypersphere/hyperboloid with radius $1/\sqrt{\beta}$. 
This geometric construction suggests that these commutation relations can be extended to be relativistically covariant. 
Furthermore, the geometry also makes the relations amenable to the introduction of gauge fields \cite{Chang:2016sae}.
Thus, for the sake of argument we will assume that the above deformed commutation relations,
or their close analog,
correctly reflect the presence of the minimal length in quantum gravity.

\section{Non-Local Extended Excitations of Quantum Fields}

Even if Eq.~(\ref{MLUR-CommutationRelations}) correctly describes the effect of quantum gravity at low energies,
the smallness of the deformation parameter $\beta$ suggests that the effect of the deformation may be difficult to detect.
Indeed, many papers exist in the literature which underscore this point.\footnote{%
See Ref.~\cite{Hossenfelder:2012jw} and references therein.  For our work, see Refs.~\cite{Chang:2001kn,Chang:2001bm,Benczik:2002tt,Benczik:2002px,Benczik:2005bh,Lewis:2011fg,Lewis:2014wfa,Chang:2016sae}.
}
However, all such investigations look at perturbations on the properties of localized particles within the QM framework.
We argue that the above relations actually point to the possible existence of non-local extended excitations of
\textit{quantum fields} whose size $\delta x$ grows linearly with its center-of-mass momentum $p$. 
Note that this is already suggested by Eq.~(\ref{MLUR2}) which implies $\delta x \sim \delta p$ for $\delta p > 1/\sqrt{\beta}$.\footnote{%
Realizing $\delta x\sim\delta p$ within deformed QM turns out to be non-trival \cite{Lewis:2011fg,Lewis:2014wfa}.
}
The new viewpoint is that instead of interpreting $\delta x$ and $\delta p$ as the uncertainties in the 
measured coordinate and momentum of a point particle, we interpret these to indicate the size of a field theoretic excitation
in phase space.

Our argument rests on the observation that our algebra, Eq.~(\ref{MLUR-CommutationRelations}),
reduces to the algebra of \textit{non-commutative field theory}\footnote{%
See Refs.~\cite{Douglas:2001ba,Szabo:2001kg,Grosse:2004yu} for reviews.}
 (NCFT) in the limit 
$\beta \to 0$, $L_{ij} \to \infty$ while keeping 
$4\beta L_{ij} \equiv \theta_{ij}$
fixed and constant \cite{Chang:2016sae}:
\begin{equation}
[\,\hat{p}_i,\,\hat{p}_j\,] 
\;=\; 0 
\;,\qquad
[\,\hat{x}_i,\,\hat{p}_j\,]
\;=\; i\hbar\,\delta_{ij}
\;,\qquad
[\,\hat{x}_i,\,\hat{x}_j\,]
\;=\; i\hbar\,\theta_{ij}
\;.
\end{equation}
In this limit, if $\theta_{ij}$ are considered to be $c$-numbers, 
the $\hat{p}_i$ operator can be identified with the adjoint operator
\begin{equation}
\hat{p}_i \;=\; [\,-\omega_{ij}\hat{x}_j,\,*\,]\;, 
\label{pdef-NCFT}
\end{equation}
where
\begin{equation}
\theta_{ik}\omega_{kj} \;=\;
\omega_{ik}\theta_{kj} \;=\; \delta_{ij}\;.
\end{equation}
NCFT interprets the $\hat{x}_i$ operators as spacetime points,
in a similar vein to matrix theory, from which the quantum field operators are constructed.

Note the proportionality of $\hat{p}$ and $\hat{x}$ in Eq.~(\ref{pdef-NCFT}).
As a consequence, the eigen-operators of the $\hat{p}$ operators (plane waves)
represent open-string-like dipole excitations for which $\delta x_i = \theta_{ij}p_j$, that is,
their spatial extent grows with their momentum \cite{SheikhJabbari:1999vm,Bigatti:1999iz}.
These states are similar to electron-hole pairs in a strong magnetic field $B$ for which
$[\hat{x}_i,\,\hat{x}_j] = i\epsilon_{ij}\ell_M^2$, where 
$\ell_M = \sqrt{\hbar c/eB}$ is the magnetic length.
The sizes of these non-local excitations in NCFT
are set by the length scale $\ell_\theta = \sqrt{\hbar\theta_{ij}}$.

Given that the NCFT commutation relation is realized in sectors with large constant angular momentum,
the theory defined by Eq.~(\ref{MLUR-CommutationRelations}) should include such non-local
extended excitations as well. 
Furthermore, since $\theta_{ij} = 4\beta L_{ij}$ in our approach, $\ell_\theta=2\sqrt{\hbar\beta L_{ij}}$ can be made as large 
as one likes without breaking rotational invariance by going to sectors of large angular momentum $L_{ij}$.
In particular, $\ell_\theta$ can be made much larger than the Planck length $\ell_P$.
Thus, the possibility exists that some of these string-like excitations are actually macroscopic, 
or even astronomical in size.
The existence of these non-particle quanta which spread with the increase of energy should be considered the new ``smoking-gun'' of quantum gravity. 
The obvious question is, how would one observe such excitations?

\section{Dynamical Phase Space}

Before addressing the above question, let us make a few more observations.

First, the above discussion points to the necessity of interpreting the 
operators in Eq.~(\ref{MLUR-CommutationRelations}) as representing points in a universal phase space,
not as the coordinate and momentum of one particular particle, and of 
constructing a field theory on the full phase space so that NCFT would
be realized on a particular ``slice.''
This may not be as wild an idea as it may seem since NCFT itself
is already a field theory constructed on a ``phase space,''
the relation $[\hat{x}_i,\hat{x}_j]=i\hbar\theta_{ij}$ essentially being the commutation relation 
between conjugate variables.\footnote{%
Formulating QM in full phase space has been discussed, for instance, in Ref.~\cite{Zachos:2001ux}.
}\footnote{%
A complete formulation of quantum field theory on phase space may also shed light on the connection
between quantum and classical theory via the Koopman-von Neumann-Sudarshan formulation of classical mechanics \cite{Koopman:1931,vNeumann1:1932,vNeumann2:1932,Sudarshan:1976bt}.
}

Second, if proportionality between momentum and coordinate as in Eq.~(\ref{pdef-NCFT}) is to hold
in particular sectors of the theory, the curved spacetime of gravity would immediately imply
a curved and dynamical energy-momentum space.
This also suggests that Eq.~(\ref{MLUR-CommutationRelations}) should be generalized to allow
for non-commuting momenta.
Again, though speculative,
this type of extension may be what is necessary for the quantization of gravity,
and the solution of the cosmological constant problem\footnote{%
We note that these points are also implied by the framework of metastring theory \cite{Freidel:2013zga,Freidel:2014qna,Freidel:2015pka,Freidel:2015uug}.
}
\cite{Chang:2010ir}.

\section{Extended Excitations as Dark Matter}

Let us now consider where we should be looking for the non-local excitations.
One obvious possibility is that these excitations were produced copiously in the early universe and 
they are still around as dark matter.
Due to their extended nature in both space and momentum, once created, it may be difficult for such excitations to decay
into local ones which are localized in either space or momentum.\footnote{%
In a model of spacetime non-locality proposed in Ref.~\cite{Saravani:2015rva},
non-local excitations, once created, cannot decay back into local excitations at all.
}
That is not to say that they cannot decay.  
The dipoles that appear in NCFT can decay into smaller dipoles, and eventually break into more localized quanta.
The speed of their decay will most probably depend on their sizes, coupling strengths, and the ease in which energy-momentum can be conserved \cite{Eby:2015hyx}.

Universality of gravity also suggests that all quantum fields, including those in the SM, should have extended non-local excitations of the type discussed here, in addition to the standard particle-like excitations.
However, since SM excitations are all charged, their presence, with the exception of the neutrino, would probably be detectable
via coherent interactions with photons or some other SM quanta.  
Thus, those would not be ``dark'' and would have had to decay early on in the history of the universe,
perhaps even before the universe became transparent.

Some weakly interacting chargeless non-local quanta would be the perfect candidates for dark matter.
These could be the non-local versions of neutrinos, or some non-SM fields yet to be detected.
Their non-locality would explain their invisibility to dark-matter detection experiments which rely
on localized interactions between localized quanta.
Their extended nature could also render them sensitive to global aspects of spacetime, such as
the Hubble parameter, and may explain Milgram scaling observed in spiral galaxies\cite{Milgrom:1983ca,Milgrom:1983pn,Milgrom:1983zz}.\footnote{%
See Refs.~\cite{Ho:2010ca,Ho:2011xc,Ho:2012ar,Edmonds:2013hba,Edmonds:2016tio} for our attempt to construct
a dark matter model of this type.
}

The non-localized excitations are highly energetic due to their large $\delta p\sim\delta x$.
So the localized daughter excitations that emerge from their decay can be expected to be ``hot''
and escape the local cold-dark-matter (CDM) halo.
So if the CDM halos around galaxies consist of non-local excitations, 
we expect their density to decrease with time.
In fact, this effect may have already been observed in elliptical galaxies \cite{Ciardullo:2002ja,Romanowsky:2003qv,Dekel:2005zn}.
The expected dependence of the speed of the decay to the size of the non-local excitation also suggests that
CDM halos of different sizes may have different decay times.
This may explain the apparent absence of dark matter in globular clusters.\footnote{%
Though it is often stated that globular clusters lack dark matter halos, this issue seems to be far from settled observationally.
See Refs.~\cite{Mashchenko:2004hj,Mashchenko:2004hk,Conroy:2010bs,Taylor:2015kka}.
}

Thus, the main message of this essay is: quantum gravity predicts the existence of non-local excitations
which could constitute the CDM halos seen in astronomical observations, and the mass density of such CDM halos
would be both size and time dependent.

To summarize, in this essay we have argued that the two generic features of quantum gravity are the existence of extended, non-local, non-particle-like quanta accompanied with a curved and, in general, dynamical energy-momentum space.
Our main point is that the phenomenology of quantum gravity does not have to be associated with the Planckian energy scale.
These general effects of quantum gravity should be found in the low energy contexts of dark matter \cite{Ho:2010ca,Ho:2011xc,Ho:2012ar,Edmonds:2013hba,Edmonds:2016tio} and dark energy \cite{Chang:2010ir}. 
Also the actual existence of the minimal length might be established using LIGO-like laser interferometry \cite{Chen:2013sgs}. 
Thus, the coming years of exploration of available energy scales should open up new and exciting avenues in the quest for the observable physical effects of quantum gravity.

\vskip 0.5cm

\section*{Acknowledgments:} 
We thank Laurent Freidel and Rob Leigh for numerous insightful discussions over many years on the topic of quantum gravity. 
We also thank Duncan Farrah and Massimo Stiavelli for informative discussions on dark matter.
The work of DM is supported in part by the U.S. Department of Energy, grant DE-FG02-13ER41917, Task A.

\bibliographystyle{ws-ijmpd}
\bibliography{MLUR}

\begin{thebibliography}{10}

\bibitem{Abbott:2016blz}
 Virgo, LIGO Scientific Collaboration (B.~P. Abbott {\em et~al.}), {\em Phys.
  Rev. Lett.} {\bf 116}  (2016)   061102,
  \href{http://arxiv.org/abs/1602.03837}{{\ttfamily arXiv:1602.03837 [gr-qc]}}.

\bibitem{Aad:2012tfa}
 ATLAS Collaboration (G.~Aad {\em et~al.}), {\em Phys. Lett.} {\bf B716}
  (2012) 1, \href{http://arxiv.org/abs/1207.7214}{{\ttfamily arXiv:1207.7214
  [hep-ex]}}.

\bibitem{Chatrchyan:2012xdj}
 CMS Collaboration (S.~Chatrchyan {\em et~al.}), {\em Phys. Lett.} {\bf B716}
  (2012) 30, \href{http://arxiv.org/abs/1207.7235}{{\ttfamily arXiv:1207.7235
  [hep-ex]}}.

\bibitem{Kiefer:2012boa}
C.~Kiefer, {\em {Quantum gravity}}, International series of monographs on
  physics, Vol.~155 (Oxford University Press, 2012).

\bibitem{Garay:1994en}
L.~J. Garay, {\em Int. J. Mod. Phys.} {\bf A10}  (1995) 145,
  \href{http://arxiv.org/abs/gr-qc/9403008}{{\ttfamily arXiv:gr-qc/9403008
  [gr-qc]}}.

\bibitem{Chang:2011jj}
L.~N. Chang, Z.~Lewis, D.~Minic and T.~Takeuchi, {\em Adv. High Energy Phys.}
  {\bf 2011}  (2011)   493514, \href{http://arxiv.org/abs/1106.0068}{{\ttfamily
  arXiv:1106.0068 [hep-th]}}.

\bibitem{Hossenfelder:2012jw}
S.~Hossenfelder, {\em Living Rev. Rel.} {\bf 16}  (2013)  ~2,
  \href{http://arxiv.org/abs/1203.6191}{{\ttfamily arXiv:1203.6191 [gr-qc]}}.

\bibitem{Chen:2013sgs}
Y.~Chen, {\em J. Phys.} {\bf B46}  (2013)   104001,
  \href{http://arxiv.org/abs/1302.1924}{{\ttfamily arXiv:1302.1924
  [quant-ph]}}.

\bibitem{Chang:2016sae}
L.~N. Chang, D.~Minic, A.~Roman, C.~Sun and T.~Takeuchi, { {On the Physics of
  the Minimal Length: The Question of Gauge Invariance}}, in {\em
  {International Conference on 60 Years of Yang-Mills Gauge Field Theories
  Singapore, Singapore, May 25-28, 2015}\/},  (2016).
\newblock \href{http://arxiv.org/abs/1602.07752}{{\ttfamily arXiv:1602.07752
  [hep-th]}}.

\bibitem{Chang:2001kn}
L.~N. Chang, D.~Minic, N.~Okamura and T.~Takeuchi, {\em Phys. Rev.} {\bf D65}
  (2002)   125027, \href{http://arxiv.org/abs/hep-th/0111181}{{\ttfamily
  arXiv:hep-th/0111181 [hep-th]}}.

\bibitem{Chang:2001bm}
L.~N. Chang, D.~Minic, N.~Okamura and T.~Takeuchi, {\em Phys. Rev.} {\bf D65}
  (2002)   125028, \href{http://arxiv.org/abs/hep-th/0201017}{{\ttfamily
  arXiv:hep-th/0201017 [hep-th]}}.

\bibitem{Benczik:2002tt}
S.~Benczik, L.~N. Chang, D.~Minic, N.~Okamura, S.~Rayyan and T.~Takeuchi, {\em
  Phys. Rev.} {\bf D66}  (2002)   026003,
  \href{http://arxiv.org/abs/hep-th/0204049}{{\ttfamily arXiv:hep-th/0204049
  [hep-th]}}.

\bibitem{Benczik:2002px}
S.~Benczik, L.~N. Chang, D.~Minic, N.~Okamura, S.~Rayyan and T.~Takeuchi
  (2002) \href{http://arxiv.org/abs/hep-th/0209119}{{\ttfamily
  arXiv:hep-th/0209119 [hep-th]}}.

\bibitem{Benczik:2005bh}
S.~Benczik, L.~N. Chang, D.~Minic and T.~Takeuchi, {\em Phys. Rev.} {\bf A72}
  (2005)   012104, \href{http://arxiv.org/abs/hep-th/0502222}{{\ttfamily
  arXiv:hep-th/0502222 [hep-th]}}.

\bibitem{Lewis:2011fg}
Z.~Lewis and T.~Takeuchi, {\em Phys. Rev.} {\bf D84}  (2011)   105029,
  \href{http://arxiv.org/abs/1109.2680}{{\ttfamily arXiv:1109.2680 [hep-th]}}.

\bibitem{Lewis:2014wfa}
Z.~Lewis, A.~Roman and T.~Takeuchi, {\em Int. J. Mod. Phys.} {\bf A30}  (2015)
   1550206, \href{http://arxiv.org/abs/1402.7191}{{\ttfamily arXiv:1402.7191
  [hep-th]}}.

\bibitem{Douglas:2001ba}
M.~R. Douglas and N.~A. Nekrasov, {\em Rev. Mod. Phys.} {\bf 73}  (2001) 977,
  \href{http://arxiv.org/abs/hep-th/0106048}{{\ttfamily arXiv:hep-th/0106048
  [hep-th]}}.

\bibitem{Szabo:2001kg}
R.~J. Szabo, {\em Phys. Rept.} {\bf 378}  (2003) 207,
  \href{http://arxiv.org/abs/hep-th/0109162}{{\ttfamily arXiv:hep-th/0109162
  [hep-th]}}.

\bibitem{Grosse:2004yu}
H.~Grosse and R.~Wulkenhaar, {\em Commun. Math. Phys.} {\bf 256}  (2005) 305,
  \href{http://arxiv.org/abs/hep-th/0401128}{{\ttfamily arXiv:hep-th/0401128
  [hep-th]}}.

\bibitem{SheikhJabbari:1999vm}
M.~M. Sheikh-Jabbari, {\em Phys. Lett.} {\bf B455}  (1999) 129,
  \href{http://arxiv.org/abs/hep-th/9901080}{{\ttfamily arXiv:hep-th/9901080
  [hep-th]}}.

\bibitem{Bigatti:1999iz}
D.~Bigatti and L.~Susskind, {\em Phys. Rev.} {\bf D62}  (2000)   066004,
  \href{http://arxiv.org/abs/hep-th/9908056}{{\ttfamily arXiv:hep-th/9908056
  [hep-th]}}.

\bibitem{Zachos:2001ux}
C.~K. Zachos, {\em Int. J. Mod. Phys.} {\bf A17}  (2002) 297,
  \href{http://arxiv.org/abs/hep-th/0110114}{{\ttfamily arXiv:hep-th/0110114
  [hep-th]}}.

\bibitem{Koopman:1931}
B.~O. Koopman, {\em Proceedings of the National Academy of Sciences of the USA}
  {\bf 17}  (1931) 315.

\bibitem{vNeumann1:1932}
J.~von Neumann, {\em Annals of Mathematics} {\bf 33}  (1932) 587.

\bibitem{vNeumann2:1932}
J.~von Neumann, {\em Annals of Mathematics} {\bf 33}  (1932) 789.

\bibitem{Sudarshan:1976bt}
G.~Sudarshan, {\em Pramana} {\bf 6}  (1976)   117.

\bibitem{Freidel:2013zga}
L.~Freidel, R.~G. Leigh and D.~Minic, {\em Phys. Lett.} {\bf B730}  (2014) 302,
  \href{http://arxiv.org/abs/1307.7080}{{\ttfamily arXiv:1307.7080 [hep-th]}}.

\bibitem{Freidel:2014qna}
L.~Freidel, R.~G. Leigh and D.~Minic, {\em Int. J. Mod. Phys.} {\bf D23}
  (2014)   1442006, \href{http://arxiv.org/abs/1405.3949}{{\ttfamily
  arXiv:1405.3949 [hep-th]}}.

\bibitem{Freidel:2015pka}
L.~Freidel, R.~G. Leigh and D.~Minic, {\em JHEP} {\bf 06}  (2015)   006,
  \href{http://arxiv.org/abs/1502.08005}{{\ttfamily arXiv:1502.08005
  [hep-th]}}.

\bibitem{Freidel:2015uug}
L.~Freidel, R.~G. Leigh and D.~Minic, {\em Int. J. Mod. Phys.} {\bf D24}
  (2015)   1544028.

\bibitem{Chang:2010ir}
L.~N. Chang, D.~Minic and T.~Takeuchi, {\em Mod. Phys. Lett.} {\bf A25}  (2010)
  2947, \href{http://arxiv.org/abs/1004.4220}{{\ttfamily arXiv:1004.4220
  [hep-th]}}.

\bibitem{Saravani:2015rva}
M.~Saravani and S.~Aslanbeigi, {\em Phys. Rev.} {\bf D92}  (2015)   103504,
  \href{http://arxiv.org/abs/1502.01655}{{\ttfamily arXiv:1502.01655
  [hep-th]}}.

\bibitem{Eby:2015hyx}
J.~Eby, P.~Suranyi and L.~C.~R. Wijewardhana  (2015)
  \href{http://arxiv.org/abs/1512.01709}{{\ttfamily arXiv:1512.01709
  [hep-ph]}}.

\bibitem{Milgrom:1983ca}
M.~Milgrom, {\em Astrophys. J.} {\bf 270}  (1983) 365.

\bibitem{Milgrom:1983pn}
M.~Milgrom, {\em Astrophys. J.} {\bf 270}  (1983) 371.

\bibitem{Milgrom:1983zz}
M.~Milgrom, {\em Astrophys. J.} {\bf 270}  (1983) 384.

\bibitem{Ho:2010ca}
C.~M. Ho, D.~Minic and Y.~J. Ng, {\em Phys. Lett.} {\bf B693}  (2010) 567,
  \href{http://arxiv.org/abs/1005.3537}{{\ttfamily arXiv:1005.3537 [hep-th]}}.

\bibitem{Ho:2011xc}
C.~M. Ho, D.~Minic and Y.~J. Ng, {\em Gen. Rel. Grav.} {\bf 43}  (2011) 2567,
  \href{http://arxiv.org/abs/1105.2916}{{\ttfamily arXiv:1105.2916 [gr-qc]}},
  [Int. J. Mod. Phys.D20,2887(2011)].

\bibitem{Ho:2012ar}
C.~M. Ho, D.~Minic and Y.~J. Ng, {\em Phys. Rev.} {\bf D85}  (2012)   104033,
  \href{http://arxiv.org/abs/1201.2365}{{\ttfamily arXiv:1201.2365 [hep-th]}}.

\bibitem{Edmonds:2013hba}
D.~Edmonds, D.~Farrah, C.~M. Ho, D.~Minic, Y.~J. Ng and T.~Takeuchi, {\em
  Astrophys. J.} {\bf 793}  (2014)  ~41,
  \href{http://arxiv.org/abs/1308.3252}{{\ttfamily arXiv:1308.3252
  [astro-ph.CO]}}.

\bibitem{Edmonds:2016tio}
D.~Edmonds, D.~Farrah, C.~m. Ho, D.~Minic, Y.~J. Ng and T.~Takeuchi  (2016)
  \href{http://arxiv.org/abs/1601.00662}{{\ttfamily arXiv:1601.00662
  [astro-ph.CO]}}.

\bibitem{Ciardullo:2002ja}
R.~Ciardullo, G.~H. Jacoby and H.~B. Dejonghe, {\em Astrophys. J.} {\bf 414}
  (1993) 454.

\bibitem{Romanowsky:2003qv}
A.~J. Romanowsky, N.~D. Douglas, M.~Arnaboldi, K.~Kuijken, M.~R. Merrifield,
  N.~R. Napolitano, M.~Capaccioli and K.~C. Freeman, {\em Science} {\bf 301}
  (2003) 1696, \href{http://arxiv.org/abs/astro-ph/0308518}{{\ttfamily
  arXiv:astro-ph/0308518 [astro-ph]}}.

\bibitem{Dekel:2005zn}
A.~Dekel, F.~Stoehr, G.~A. Mamon, T.~J. Cox and J.~R. Primack, {\em Nature}
  {\bf 437}  (2005)   707,
  \href{http://arxiv.org/abs/astro-ph/0501622}{{\ttfamily
  arXiv:astro-ph/0501622 [astro-ph]}}.

\bibitem{Mashchenko:2004hj}
S.~Mashchenko and A.~Sills, {\em Astrophys. J.} {\bf 619}  (2005)   243,
  \href{http://arxiv.org/abs/astro-ph/0409605}{{\ttfamily
  arXiv:astro-ph/0409605 [astro-ph]}}.

\bibitem{Mashchenko:2004hk}
S.~Mashchenko and A.~Sills, {\em Astrophys. J.} {\bf 619}  (2005)   258,
  \href{http://arxiv.org/abs/astro-ph/0409606}{{\ttfamily
  arXiv:astro-ph/0409606 [astro-ph]}}.

\bibitem{Conroy:2010bs}
C.~Conroy, A.~Loeb and D.~Spergel, {\em Astrophys. J.} {\bf 741}  (2011)  ~72,
  \href{http://arxiv.org/abs/1010.5783}{{\ttfamily arXiv:1010.5783
  [astro-ph.GA]}}.

\bibitem{Taylor:2015kka}
M.~A. Taylor, T.~H. Puzia, M.~Gomez and K.~A. Woodley, {\em Astrophys. J.} {\bf
  805}  (2015)  ~65.

\end{thebibliography}

\end{document}